# Unidirectional Cyclic Quantum Teleportation of Arbitrary Schrodinger Cat Coherent-states via Bell Coherent-States

Ankita Pathak* and Ravi S. Singh†

Photonic Quantum-Information and Quantum Optics Group, Department of Physics,

Deen Dayal Upadhyaya Gorakhpur University, Gorakhpur (U.P.), 273009, India.

e.mail: †yesora27@gmail.com ; *pathak18.phy@gmail.com

## Abstract

Exploiting the cluster of three Bell coherent-states as quantum channel, we presented a scheme wherein quantum-informations encoded in three arbitrary superposed coherent states, i.e., Schrodinger Cat coherent-states are simultaneously transmitted in unidirectional cyclic sequence among three parties by invoking usage of linear optical gadgets such as ideal beam-splitters, phase shifters and photon-number-resolving detectors. Proposed simultaneous unidirectional Cyclic Quantum Teleportation is faithful with one-eighth probability of success. Furthermore, it is seen that not every detection-events provides faithful scheme and, hence, fidelity and probability of success of near-faithful Cyclic Quantum Teleportation is evaluated and numerically interpreted.

## I. Introduction

Quantum Teleportation (QT) is a counterintuitive communication process wherein transmission of quantum information from one location to another location is discovered by Bennett et. al. [1] in a theoretical protocol via employing Bell-pair [2] as the quantum channel without physically transferring quantum particles preserving quantum-information. This scheme is, popularly, termed as 'Standard Quantum Teleportation' (SQT) in the discrete-variable (DV) regime. SQT is extended to continuous-variable (CV) regime by Braunstein and Kimble [3] in which a strategy has been put forth to teleport the wave function of single mode electromagnetic field using bi-modal vacuum squeezed state.

Notably, SQT is one of the myriad communication tasks such as quantum cryptography [4-7], super dense coding [8-10], quantum secure direct communication [11-16], quantum secret sharing [17-19], remote state preparation [20-23], port-based QT [24-26], counterfactual quantum communication [27-29], Catalytic QT [30-33]. SQT is realized experimentally by using variant physical systems such as photonic states [34-39], optical quantum modes [40-42], nuclear magnetic resonance [43], atomic ensembles states [44-46], trapped atoms states [47-49] or solid-state systems' states [50-52], which are furnishing quantum technologies in demonstrating quantum information processing tasks. Moreover, QT is being conjoined with variety of concepts to put forward novel protocols such as quantum telecomputing [53-55], quantum telecloning [56-58], quantum telefilters and telemirrors [59], quantum broadcasting [60-61], quantum tele- amplification [62], quantum digital signature[63] to name a few.

SQT in DV regime is generalized by Karlsson and Bourennane [64], realized in experiments by Barasinski et. al. and Liu et. al. [65-66], wherein a scheme is proposed involving a third party intertwined via GHZ states [67] as quantum channel. The third-party controls perfect reception of quantum state at receiver's end and, hence, the name 'controlled QT' is accorded. Controlled QT via W-state [68-71] and GHZ-like-states [72] as quantum channel is also investigated. Unidirectional controlled QT has outgrown into many variants via involving many-parties and multi-qubits and addressing security concerns [73-75].

An alternative procedure for encoding quantum information by even and odd coherent states [76], historically recognized as Schrodinger Cat coherent-states, which spanned finite two-dimensional Hilbert space although individual coherent states lied in unbounded infinite dimensional Hilbert space, is introduced by Cochrane et. al.[77]. Soon after, Ralph et. al. [78-80] opined that the logical qubits, $\{|0\rangle,|1\rangle\}$ may be associated with optical coherent states, $\{|\alpha\rangle,|-\alpha\rangle\}$, distinguished by out of phase $\pi$ and regarded as two non-orthogonal linearly independent multiphoton states, which may be used to encode quantum information leading to investigation for universal quantum computation. Using ideal linear optical devices such as beam splitters, phase shifters, and photon number-resolving detectors van Enk and Hirota [81] devised a scheme for teleporting a Schrodinger Cat coherent-states via bi-modal entangled coherent states [82-83], a non-Gaussian optical quantum field possessing higher-order optical polarization [84], as the quantum channel. Wang [85] extended the protocol [81] for transmitting bi-modal Bell-entangled coherent states using tri-modal GHZ-coherent states as quantum channel, which is further generalized for teleporting arbitrary bi-modal coherent states via schemes put forward by Liao and Kuang [86], Phien and An [87], Prakash and Mishra [88].

Controlled QT of Schrodinger Cat- coherent States is worked out via GHZ-coherent states as the quantum channel [89]. A symmetric bi-directional controlled QT of Schrodinger Cat coherent-states using penta-modal cluster-type entangled coherent states is chalked out by Pandey et. al. [90], of which counterpart in DV regime is proposed by Zha et. al. and Shukla et. al. [91-92] and experimentally realized by Grebel et. al. [93]. Moreover, asymmetric bi-directional exchange of bi-modal Bell-coherent states and Schrodinger Cat-coherent states is worked out via a hexa-modal entangled coherent state consisting of tensor product of one GHZ- coherent states and one Bell-coherent states [94]. Recently, Chen et. al. [95] investigated first theoretical protocol wherein three parties are involved in transmission of three arbitrary single qubits in the cyclic sequence and, therefore, the term Cyclic QT is coined. Following the lineage of Chen et. al. [95] and criticizing bidirectional controlled quantum teleportation involving three parties put forth by Zhang [96], Verma [97] proposed a general protocol for Cyclic QT.

Cyclic QT neither in CV regime nor in schemes utilizing optical coherent states, an indispensable and imperative need in transmitting quantum information within a quantum network, has not been witnessed in literature. It is, therefore, we came up with a strategy involving three parties for transmitting three arbitrary Schrodinger Cat coherent- states in unidirectional cyclic sequence via employing a cluster of three Bell coherent- states as quantum channel. The write-up is structured in following Sections. Section (II) describes detailed protocol for Cyclic QT. Section (III) deals with the probability of success and fidelity of faithful unidirectional cyclic QT. Section (IV) is devoted to analyze near-faithful partial unidirectional cyclic QT. Finally, conclusions and future perspectives are delineated in Section IV.

## II. Scheme for Simultaneous Unidirectional Teleportation in Cyclic Sequence

The quantum channel shared amongst communicating-parties (Alice, Bob and Charlie) consist of three Bell states,

$$|\psi\rangle_{1,2,3,4,5,6} = N_{Ch}\left(|\Omega\rangle_{1,4} \otimes |\Omega\rangle_{2,5} \otimes |\Omega\rangle_{3,6}\right) \tag{1}$$

where, $|\Omega\rangle$=$N(|\alpha,\alpha\rangle + |-\alpha,-\alpha\rangle)$, and $N_{Ch} = \left[8\left(1 + e^{-12|\alpha|^2} + 3e^{-8|\alpha|^2} + 3e^{-4|\alpha|^2}\right)\right]^{-1/2}$ is normalization constant. Obviously, one may prepare Bell Coherent-states, say, $|\Omega\rangle_{\,i,j}$ as $|\Omega\rangle_{i,j} = \hat{B}\left(\tilde{N}|\alpha\sqrt{2}\rangle + |-\alpha\sqrt{2}\rangle\right)_u \otimes |0\rangle_v = \tilde{N}\left(|\alpha,\alpha\rangle + |-\alpha,-\alpha\rangle\right)_{i,j}$ ,where $\hat{B}$ corresponds to "symmetric beam splitter with phase shifter" (BPS) operation, such that

$$\hat{\mathrm{B}}\left(|\alpha\rangle_u \otimes |\beta\rangle_v\right) = \left|\frac{\alpha+\beta}{\sqrt{2}}\right\rangle_i, \left|\frac{\alpha-\beta}{\sqrt{2}}\right\rangle_j \tag{2}$$

The quantum-informations possessed by Alice, Bob and Charlie, are encoded in even Schrodinger Cat Coherent- states [76],

$$|\psi\rangle_a = N_a(a_0|\alpha\rangle + a_1|-\alpha\rangle), \tag{3}$$

$$|\psi\rangle_b = N_b(b_0|\alpha\rangle + b_1|-\alpha\rangle), \tag{4}$$

$$|\psi\rangle_c = N_c(c_0|\alpha\rangle + c_1|-\alpha\rangle) \tag{5}$$

where, nnormalization constants are, $N_a = |a_0|^2 + |a_1|^2 + 2e^{-4|\alpha|^2} Re\,(a_0,a_1^*)$, $N_b = |b_0|^2 + |b_1|^2 + 2e^{-4|\alpha|^2} Re\,(b_0,b_1^*)$, $N_c = |c_0|^2 + |c_1|^2 + 2e^{-4|\alpha|^2} Re\,(c_0,c_1^*)$ and $Re$ stands for 'real part'.

QT in cyclic sequence is initiated by assigning modes of entangled channel, Eq. (1), with following distribution-strategy: modes (1, 6) to Alice, modes (2, 4) to Bob and modes (3, 5) to Charlie. The composite system of entire modes may, obviously, be described by the global state,

$$|\phi\rangle_{a,b,c,1,2,3,4,5,6} = N_T\left(|\psi\rangle_a \otimes |\psi\rangle_b \otimes |\psi\rangle_c \otimes |\psi\rangle_{1,2,3,4,5,6}\right),$$

where, $N_T = N \times N_{ch}$ and $N = N_a \otimes N_b \otimes N_c$ is defined as the normalization constant. Inserting Eqs. (3-5), one obtains,

$$|\phi\rangle_{a,b,c,1,2,3,4,5,6} = N_T(a_0b_0c_0|\alpha,\alpha,\alpha\rangle + a_0b_0c_1|\alpha,\alpha,-\alpha\rangle + a_0b_1c_0|\alpha,-\alpha,\alpha\rangle + a_0b_1c_1|\alpha,-\alpha,-\alpha\rangle + a_1b_0c_0|-\alpha,\alpha,\alpha\rangle + a_1b_0c_1|-\alpha,\alpha,-\alpha\rangle + a_1b_1c_0|-\alpha,-\alpha,\alpha\rangle + a_1b_1c_1|-\alpha,-\alpha,-\alpha\rangle)_{a,b,c} \otimes |\psi\rangle_{123456} \quad (6)$$

Simultaneous unidirectional Cyclic QT protocol may be outlined in following steps:

**Step-1**: Alice's information-state 'a' is mixed with mode'1' of entangled channel using symmetric beam splitter in-built with phase shifters (BPS-1) of which output modes are 7 and 8. Similarly, Bob's information-state 'b' is combined with mode '2' of entangled channel via BPS-2 having output modes 9,10 and Charlie mixes information-state 'c' with mode '3' of entangled channel via BPS-3 with output modes 11 and 12 (see Figure-1).

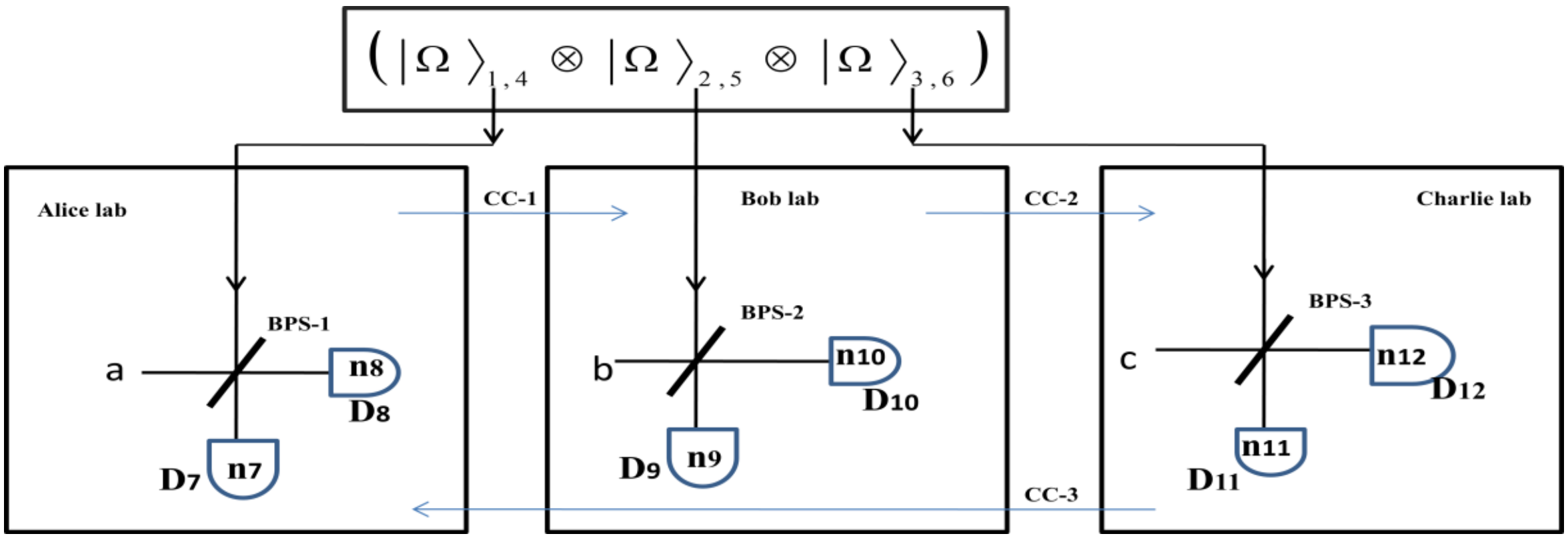

Figure1: Schematic diagram of the proposed scheme for simultaneous unidirectional Cyclic QT. Alice mixes mode (a,1), Bob mixes mode (b,2) and Charlie mixes mode (c,3) using (BPS 1-3), respectively, and they performed photon counting measurement on output modes 7-12 using photon number resolving detector detectors $D_{7-12}$. and communicate their counts, $n_{7-12}$ to Bob, Charlie and back to Alice using classical channels ( →).

Applying Eq. (2) along with modes-mixing procedure described in Step-1, Figure.1, we obtain the global state, Eq. (A.1), in Appendix A. Eq. (A.1), clearly, displays the presence of null (vacuum) photons at various output-modes which play an important role (see Step-3, Eqs. (7a-7h) below).

**Step2**- Alice, Bob and Charlie performs photon-count measurements in their respective labs by the photon-number resolving detectors $D_{7-12}$ , which would project the global state vector, Eq. (A1) in Appendix A to a heralded state vector,

$$|\chi(\mathrm{n}_7,\mathrm{n}_8,\mathrm{n}_9,\mathrm{n}_{10},\mathrm{n}_{11},\mathrm{n}_{12})\rangle_{4,5,6} = \langle \mathrm{n}_7,\mathrm{n}_8,\mathrm{n}_9,\mathrm{n}_{10},\mathrm{n}_{11},\mathrm{n}_{12}|\psi\rangle_{7,8,9,10,11,12,4,5,6} . \tag{7}$$

Now, Alice sends photon-counts, $n_7, n_8$ to Bob, Bob do the same, $n_9, n_{10}$ to Charlie and Charlie communicates his photon-counts, $n_{11}, n_{12}$ to Alice, simultaneously, via classical channels.

***Step*** **3**-Alice, Bob and Charlie may find non-zero and null photon-counts at their respective output ports, i.e., one may obtain photon-counts as $n_7 \neq 0, n_8 = 0$ from detectors $D_{7,8}$; photons $n_9 \neq 0, n_{10} = 0$, from detector $D_{9,10}$, and photons $n_{11} \neq 0, n_{12} = 0$ from detectors $D_{11,12}$. Obviously, eight such cases would arise.

**Case-I :** For photon-counts, $n_7 \neq 0, n_8 = 0, n_9 \neq 0, n_{10} = 0,\ n_{11} \neq 0, n_{12} = 0$, a heralded state

$$\begin{aligned}|\chi^1\rangle_{4,5,6} &= \langle \mathrm{n}_7,0,\mathrm{n}_9,0,\mathrm{n}_{11},0|\psi\rangle_{7,8,9,10,11,12,4,5,6} = N_T\,(a_0b_0c_0|\alpha,\alpha,\alpha\rangle + (-1)^{n_{11}}\,a_0b_0c_1|\alpha,\alpha,-\alpha\rangle + \\ &(-1)^{n_9}\,a_0b_1c_0|\alpha,-\alpha,\alpha\rangle + (-1)^{n_9+n_{11}}\,a_0b_1c_1|\alpha,-\alpha,-\alpha\rangle + (-1)^{n_7}\,a_1b_0c_0|-\alpha,\alpha,\alpha\rangle + (-1)^{n_7+n_{11}} \\ &a_1b_0c_1|-\alpha,\alpha,-\alpha\rangle + (-1)^{n_7+n_9}\,a_1b_1c_0|-\alpha,-\alpha,\alpha\rangle + (-1)^{n_7+n_9+n_{11}}\,a_1b_1c_1|-\alpha,-\alpha,-\alpha\rangle)_{4,5,6}\end{aligned} \tag{7a}$$

would result. Evidently, if $n_7, n_9\ and\ n_{11}$ are even, the perfect or faithful (unit fidelity) simultaneous unidirectional Cyclic QT is accomplished for Eq. (7a) yields, $|\chi^1\rangle_{4,5,6} = N(a_o|\alpha\rangle + a_1|-\alpha\rangle)_4 \otimes (b_o|\alpha\rangle + b_1|-\alpha\rangle)_5 \otimes (c_o|\alpha\rangle + c_1|-\alpha\rangle)_6$, exact replica of $|\psi\rangle_{4,5,6}$.

One evaluates fidelity, i.e, the overlapping of 'to be teleported' information-state and the states obtained after Cyclic QT gets completed. Fidelity is, therefore, seen to have unit value.

**Case-II :** For photon-counts, $n_7 = 0, n_8 \neq 0, n_9 = 0, n_{10} \neq 0, n_{11} = 0, n_{12} \neq 0$, a heralded state

$$\begin{aligned}\left|\chi^2\right\rangle_{4,5,6} &= \langle 0, n_8, 0, n_{10}, 0, n_{12} |\psi\rangle_{7,8,9,10,11,12,4,5,6} = N_T(a_0b_0c_0|-\alpha,-\alpha,-\alpha\rangle + (-1)^{n_{12}} a_0b_0c_1|-\alpha,-\alpha,\alpha\rangle + \\ &(-1)^{n_{10}} a_0b_1c_0|-\alpha,\alpha,-\alpha\rangle + (-1)^{n_{10}+n_{12}} a_0b_1c_1|-\alpha,\alpha,\alpha\rangle + (-1)^{n_8} a_1b_0c_0|\alpha,-\alpha,-\alpha\rangle + (-1)^{n_8+n_{12}} \\ &a_1b_0c_1|\alpha,-\alpha,\alpha\rangle + (-1)^{n_8+n_{10}} a_1b_1c_0|\alpha,\alpha,-\alpha\rangle + (-1)^{n_8+n_{10}+n_{12}} a_1b_1c_1|\alpha,\alpha,\alpha\rangle)_{4,5,6}\end{aligned} \quad (7b)$$

would result. A keen look into Eq. (7b) suggests that if photon-counts, $n_8, n_{10}$ and $n_{12}$ are even, $\left|\chi^1\right\rangle_{4,5,6} = P_{4,5,6}(\theta)\left|\chi^2\right\rangle_{4,5,6}$, where, $P_i(\theta)$ is phase shifting operator for j[th]-mode, $P_j(\theta) = \exp(-ia_j^\dagger a_j)$ such that $|\alpha\rangle \to P_i(\theta)|\alpha\rangle = |e^{j\theta}\alpha\rangle$, taking $\theta = \pi$, $|\alpha\rangle = |-\alpha\rangle\, for\, j - mode$. This implies that Alice, Bob and Charlie achieve faithful simultaneous unidirectional Cyclic QT upto only phase-shifter in their respective labs.

**Case-III:** For photon-counts, $n_7 \neq 0, n_8 = 0, n_9 \neq 0, n_{10} = 0$, $n_{11} = 0, n_{12} \neq 0$, a heralded state

$$\begin{aligned}\left|\chi^3\right\rangle_{4,5,6} &= \langle n_7, 0, n_9, 0, 0, n_{12} |\psi\rangle_{7,8,9,10,11,12,4,5,6} = N_T(a_0b_0c_0|\alpha,\alpha,-\alpha\rangle + (-1)^{n_{12}} a_0b_0c_1|\alpha,\alpha,\alpha\rangle + \\ &(-1)^{n_9} a_0b_1c_0|\alpha,-\alpha,-\alpha\rangle + (-1)^{n_9+n_{12}} a_0b_1c_1|\alpha,-\alpha,\alpha\rangle + (-1)^{n_7} a_1b_0c_0|-\alpha,\alpha,-\alpha\rangle + (-1)^{n_7+n_{12}} \\ &a_1b_0c_1|-\alpha,\alpha,\alpha\rangle + (-1)^{n_7+n_9} a_1b_1c_0|-\alpha,-\alpha,-\alpha\rangle + (-1)^{n_7+n_9+n_{12}} a_1b_1c_1|-\alpha,-\alpha,\alpha\rangle)_{4,5,6}\end{aligned} \quad (7c)$$

would result. Evidently, for $n_7, n_9$ and $n_{12}$ are even and $\left|\chi^1\right\rangle_{4,5,6} = P_6(\pi)\left|\chi^3\right\rangle_{4,5,6}$, exact replica of $|\psi\rangle_{4,5,6}$, the 'to be teleported state', and, therefore perfect unidirectional Cyclic QT.

**Case-IV:** For photon-counts, $n_7 \neq 0, n_8 = 0, n_9 = 0, n_{10} \neq 0$, $n_{11} \neq 0, n_{12} = 0$, a heralded state

$$\begin{aligned}\left|\chi^4\right\rangle_{4,5,6} &= \langle n_7, 0, 0, n_{10}, n_{11}, 0|\psi\rangle_{7,8,9,10,11,12,4,5,6} = N(a_0b_0c_0\,|\alpha,-\alpha,\alpha\rangle_{4,5,6} + (-1)^{n_{11}} a_0b_0c_1\,|\alpha,-\alpha,-\alpha\rangle_{4,5,6} + \\ &(-1)^{n_{10}} a_0b_1c_0\,|\alpha,\alpha,\alpha\rangle_{4,5,6} + (-1)^{n_{10}+n_{11}} a_0b_1c_1\,|\alpha,\alpha,-\alpha\rangle_{4,5,6} + (-1)^{n_7} a_1b_0c_0\,|-\alpha,-\alpha,\alpha\rangle_{4,5,6} + \\ &+(-1)^{n_7+n_{11}} a_1b_0c_1\,|-\alpha,-\alpha,-\alpha\rangle_{4,5,6} + (-1)^{n_7+n_{10}} a_1b_1c_0\,|-\alpha,\alpha,\alpha\rangle_{4,5,6} + (-1)^{n_7+n_{10}+n_{11}} a_1b_1c_1\,|-\alpha,\alpha,-\alpha\rangle_{4,5,6})\end{aligned} \quad (7d)$$

would result. Clearly, $n_7, n_{10}$ and $n_{11}$ are even, and $\left|\chi^1\right\rangle_{4,5,6} = P_5(\pi)\left|\chi^4\right\rangle_{4,5,6}$ exact replica of $\left|\psi\right\rangle_{4,5,6}$, the 'to be teleported state', and, therefore, faithful simultaneous unidirectional Cyclic QT.

**Case-V:** For photon-counts, $n_7 \neq 0, n_8 = 0, n_9 = 0, n_{10} \neq 0, n_{11} = 0, n_{12} \neq 0$, a heralded state

$$\left|\chi^5\right\rangle_{4,5,6} = \left\langle \mathrm{n}_7, 0, 0, \mathrm{n}_{10}, 0, \mathrm{n}_{12} \right|\psi\rangle_{7,8,9,10,11,12,4,5,6} = N(a_0 b_0 c_0 \left|\alpha, -\alpha, -\alpha\right\rangle_{4,5,6} + (-1)^{n_{12}} a_0 b_0 c_1 \left|\alpha, -\alpha, \alpha\right\rangle_{4,5,6}$$
$$+ (-1)^{n_{10}} a_0 b_1 c_0 \left|\alpha, \alpha, -\alpha\right\rangle_{4,5,6} + (-1)^{n_{10}+n_{12}} a_0 b_1 c_1 \left|\alpha, \alpha, \alpha\right\rangle_{4,5,6} + (-1)^{n_7} a_1 b_0 c_0 \left|-\alpha, -\alpha, -\alpha\right\rangle_{4,5,6}$$
$$+ (-1)^{n_7+n_{12}} a_1 b_0 c_1 \left|-\alpha, -\alpha, \alpha\right\rangle_{4,5,6} + (-1)^{n_7+n_{10}} a_1 b_1 c_0 \left|-\alpha, \alpha, -\alpha\right\rangle_{4,5,6} + (-1)^{n_7+n_{10}+n_{12}} a_1 b_1 c_1 \left|-\alpha, \alpha, \alpha\right\rangle_{4,5,6}) \quad (7e)$$

would result. Obviously, if $n_7, n_{10}$ and $n_{12}$ are even, one may observe from Eq. (7e), $\left|\chi^1\right\rangle_{4,5,6} = P_{5,6}(\pi)\left|\chi^5\right\rangle_{4,5,6}$, exact replica of $\left|\psi\right\rangle_{4,5,6}$, the 'to be teleported state', and, therefore faithful simultaneous unidirectional Cyclic QT.

**Case-VI:** For photon counts $\mathrm{n}_7 = 0, \mathrm{n}_8 \neq 0, \mathrm{n}_9 \neq 0, \mathrm{n}_{10} = 0, \mathrm{n}_{11} \neq 0, \mathrm{n}_{12} = 0$, a heralded state

$$\left|\chi^6\right\rangle_{4,5,6} = \left\langle 0, \mathrm{n}_8, \mathrm{n}_9, 0, \mathrm{n}_{11}, 0 \right|\psi\rangle_{7,8,9,10,11,12,4,5,6} = N(a_0 b_0 c_0 \left|-\alpha, \alpha, \alpha\right\rangle_{4,5,6} + (-1)^{n_{11}} a_0 b_0 c_1 \left|-\alpha, \alpha, -\alpha\right\rangle_{4,5,6}$$
$$+ (-1)^{n_9} a_0 b_1 c_0 \left|-\alpha, -\alpha, \alpha\right\rangle_{4,5,6} + (-1)^{n_9+n_{11}} a_0 b_1 c_1 \left|-\alpha, -\alpha, -\alpha\right\rangle_{4,5,6} + (-1)^{n_8} a_1 b_0 c_0 \left|\alpha, \alpha, \alpha\right\rangle_{4,5,6}$$
$$+ (-1)^{n_8+n_{11}} a_1 b_0 c_1 \left|\alpha, \alpha, -\alpha\right\rangle_{4,5,6} + (-1)^{n_8+n_9} a_1 b_1 c_0 \left|\alpha, -\alpha, \alpha\right\rangle_{4,5,6} + (-1)^{n_8+n_9+n_{11}} a_1 b_1 c_1 \left|\alpha, -\alpha, -\alpha\right\rangle_{4,5,6}) \quad (7f)$$

would result. Evidently, if $n_8, n_9\, and\, n_{11}$ are even, one may see from Eq. (7f), $\left|\chi^1\right\rangle_{4,5,6} = P_4(\pi)\left|\chi^6\right\rangle_{4,5,6}$, exact replica of $\left|\psi\right\rangle_{4,5,6}$, the 'to be teleported state', and, therefore perfect simultaneous unidirectional Cyclic QT.

**Case-VII:** For photon counts $\mathrm{n}_7 = 0, \mathrm{n}_8 \neq 0, \mathrm{n}_9 \neq 0, \mathrm{n}_{10} = 0, \mathrm{n}_{11} = 0, \mathrm{n}_{12} \neq 0$, a heralded state

$$\left|\chi^7\right\rangle_{4,5,6} = \left\langle 0, \mathrm{n}_8, \mathrm{n}_9, 0, 0, \mathrm{n}_{12} \right|\psi\rangle_{7,8,9,10,11,12,4,5,6} = N_T(a_0 b_0 c_0 \left|-\alpha, \alpha, -\alpha\right\rangle_{4,5,6} + (-1)^{n_{12}} a_0 b_0 c_1 \left|-\alpha, \alpha, \alpha\right\rangle_{4,5,6}$$
$$+ (-1)^{n_9} a_0 b_1 c_0 \left|-\alpha, -\alpha, -\alpha\right\rangle_{4,5,6} + (-1)^{n_9+n_{12}} a_0 b_1 c_1 \left|-\alpha, -\alpha, \alpha\right\rangle_{4,5,6} + (-1)^{n_8} a_1 b_0 c_0 \left|\alpha, \alpha, -\alpha\right\rangle_{4,5,6}$$

$$+(-1)^{n_8+n_{12}} a_1 b_0 c_1 \left|\alpha,\alpha,\alpha\right\rangle_{4,5,6} + (-1)^{n_8+n_9} a_1 b_1 c_0 \left|\alpha,-\alpha,-\alpha\right\rangle_{4,5,6} + (-1)^{n_8+n_9+n_{12}} a_1 b_1 c_1 \left|\alpha,-\alpha,-\alpha\right\rangle_{4,5,6}) \qquad (7g)$$

would result. One may enquire that if $n_8, n_9$ and $n_{12}$ are even and noting from Eq. (7g), $\left|\chi^1\right\rangle_{4,5,6} = P_{4,6}(\pi)\left|\chi^7\right\rangle_{4,5,6}$, exact replica of $\left|\psi\right\rangle_{4,5,6}$, the 'to be teleported state', and, therefore perfect simultaneous unidirectional Cyclic QT results.

**Case-VIII:** For photon counts $n_7 = 0, n_8 \neq 0, n_9 = 0, n_{10} \neq 0, n_{11} \neq 0, n_{12} = 0,$ a heralded state,

$$\left|\chi^8\right\rangle_{4,5,6} = \left\langle 0, \mathrm{n}_8, 0, \mathrm{n}_{10}, \mathrm{n}_{11}, 0 \right|\psi\rangle_{7,8,9,10,11,12,4,5,6} = N_T(a_0 b_0 c_0 \left|-\alpha,-\alpha,\alpha\right\rangle_{4,5,6} + (-1)^{n_{11}} a_0 b_0 c_1 \left|-\alpha,-\alpha,-\alpha\right\rangle_{4,5,6}$$
$$+(-1)^{n_{10}} a_0 b_1 c_0 \left|-\alpha,\alpha,\alpha\right\rangle_{4,5,6} + (-1)^{n_{10}+n_{11}} a_0 b_1 c_1 \left|-\alpha,\alpha,-\alpha\right\rangle_{4,5,6} + (-1)^{n_8} a_1 b_0 c_0 \left|\alpha,-\alpha,\alpha\right\rangle_{4,5,6}$$
$$+(-1)^{n_8+n_{11}} a_1 b_0 c_1 \left|\alpha,-\alpha,-\alpha\right\rangle_{4,5,6} + (-1)^{n_8+n_{10}} a_1 b_1 c_0 \left|\alpha,\alpha,\alpha\right\rangle_{4,5,6} + (-1)^{n_8+n_{10}+n_{11}} a_1 b_1 c_1 \left|\alpha,\alpha,-\alpha\right\rangle_{4,5,6}) \qquad (7h)$$

would result. Obviously, if $n_8, n_{10}$ and $n_{11}$ are even and noting from Eq. (7h), $\left|\chi^1\right\rangle_{4,5,6} = P_{4,5}(\pi)\left|\chi^8\right\rangle_{4,5,6}$, exact replica of $\left|\psi\right\rangle_{4,5,6}$, the 'to be teleported state', and, therefore perfect simultaneous unidirectional Cyclic QT result.

Finally, the probability of finding the number of photons $n_{7-12}$ in detectors $D_{7-12}$ may be derived by

$$\mathrm{P}\,(\mathrm{n}_7,\mathrm{n}_8,\mathrm{n}_9,\mathrm{n}_{10},\mathrm{n}_{11},\mathrm{n}_{12}) = \sum_{\mathrm{n}_7,\mathrm{n}_8,\mathrm{n}_9,\mathrm{n}_{10},\mathrm{n}_{11},\mathrm{n}_{12}} \left|\left\langle \mathrm{n}_7,\mathrm{n}_8,\mathrm{n}_9,\mathrm{n}_{10},\mathrm{n}_{11},\mathrm{n}_{12}\right|\psi\rangle_{7,8,9,10,11,12,4,5,6}\right|^2, \qquad (8)$$

where, $\left|\psi\right\rangle_{7,8,9,10,11,12,4,5,6}$ is given in Eq. (A1) in Appendix A. Evaluation of Eq. (8) for Case-1 Eq. (7a) is $P_{\chi^1} = (n_7, n_9, n_{11}, 0,0,0) = 1/8$ for even $\mathrm{n}_8, \mathrm{n}_{10}, \mathrm{n}_{12}$ for. intense optical coherent states ($\alpha^2 \to \infty$). The total 'probability of success' for simultaneous unidirectional Cyclic QT, denoted by $P_T^{Cy}$, $P_T^{Cy} = \sum_{i=1}^{8} P_{\chi^i} = 1$. Evidently, total probability $P_T^{Cy}$ is independent of α, the modal amplitude of optical coherent field as well as the unknown probability-amplitudes $a_0$, $a_1$, $b_0$, $b_1$, and $c_0$, $c_1$ of quantum information-states

Step 4: Furthermore, many different possibilities for photon-counts, $n_{7-12}$ from detectors $D_{7-12}$ may arise, which are, in turn, to be communicated to receivers to infer corresponding unitary operations (phase-shifting or displacement operators) in Alice's, Bob's and Charlie's labs to reconstruct the states (see Tables: I-IV in Appendix B)

**III. Probability and Fidelity for Near faithful Partial Cyclic QT.**

The cases wherein at least one faithful unidirectional Cyclic QT ($A \rightarrow B\ or\ B \rightarrow C\ or\ C \rightarrow A$ ) happened, one requires the displacement operators to recover the original state, and, hence, so-called 'near-faithful' partial cyclic QT results for an intense optical field.

A close observation of Tables (I-IV) in Appendix B shows that not all detection-events accomplishes faithful Cyclic QT, e.g., Table-1 , Eq. (7a), First row (Odd, Odd, Odd).

$$|\chi^1\rangle_{4,5,6} = N_a(a_0|\alpha\rangle - a_1|-\alpha\rangle)_4 \otimes N_b(b_0|\alpha\rangle - b_1|-\alpha\rangle)_5 \otimes N_c(c_0|\alpha\rangle - c_1|-\alpha\rangle)_6 \quad (9)$$

Eq. (9), is not exactly same as information-states of Alice Bob and Charlie, but by application of displacement operator, $D_k(\delta) = \exp(\delta b_k^{\dagger} - \delta^* b_k)$, in Eq. (9), we obtain, for simplicity, parameterizing probability amplitudes by setting $a_0$= cos $\theta_1$, $a_1$= sin $\theta_1$, $b_0$= cos $\theta_2$ , $b_1$= sin $\theta_2$ , $c_0$= cos $\theta_3$ , $c_1$= sin $\theta_3$ ,

$$D_4\left(\frac{i\pi}{2\alpha}\right)|\chi^1\rangle_4 = |\zeta\rangle_4 = N_{1a}e^{i\pi/2}\left(a_0\left|\frac{i\pi}{2\alpha}+\alpha\right\rangle + a_1\left|\frac{i\pi}{2\alpha}-\alpha\right\rangle\right)_4, \quad (10)$$

$$D_5\left(\frac{i\pi}{2\alpha}\right)|\chi^1\rangle_5 = |\xi\rangle_5 = N_{1b}e^{i\pi/2}\left(b_0\left|\frac{i\pi}{2\alpha}+\alpha\right\rangle + b_1\left|\frac{i\pi}{2\alpha}-\alpha\right\rangle\right)_5, \quad (11)$$

$$D_6\left(\frac{i\pi}{2\alpha}\right)|\chi^1\rangle_6 = |\tau\rangle_6 = N_{1c}e^{i\pi/2}\left(c_0\left|\frac{i\pi}{2\alpha}+\alpha\right\rangle + c_1\left|\frac{i\pi}{2\alpha}-\alpha\right\rangle\right)_6, \quad (12)$$

Clearly, a characteristic property of displacement operator, $D_l(\beta)|\delta\rangle_l = \exp\left[\frac{1}{2}\left(\beta\delta^* - \beta^*\delta\right)\right]|\beta+\delta\rangle_l$ has been used. By virtue of scalar product of optical coherent states, $\langle\beta|\delta\rangle = \exp\left[-\left(|\beta|^2 + |\delta|^2 - 2\beta^*\delta\right)/2\right]$, the fidelity at Bob's lab $F_1^{(I)A\to B}\left|{}_a\langle\psi|\zeta\rangle_4\right|^2$, that at Charlie's lab $F_1^{(I)B\to C}\left|{}_b\langle\psi|\tau\rangle_5\right|^2$ and that Alice's lab, $F_1^{(I)C\to A}\left|{}_c\langle\psi|\omega\rangle_6\right|^2$ may be evaluated by insertion of Eq. (10-12), which are, obviously, not be equal to unity, for,

$$F_1^{(I)A\to B} = \left(N_{1a}N_a\right)^2 e^{\frac{-\pi^2}{8\alpha^2}}\left(\sum_{i=0}^{1} a_i + 2e^{-2\alpha^2}\left(a_0a_1\right)\right)^2, P_1^{(I)A\to B} = \left(\frac{N_a}{8N_{1a}}\right)^2\left(\frac{1-e^{-2\alpha^2}}{1+e^{-2\alpha^2}}\right)^2, \quad (13)$$

$$F_1^{(I)B\to C} = \left(N_{1b}N_b\right)^2 e^{\frac{-\pi^2}{8\alpha^2}}\left(\sum_{i=0}^{1} b_i + 2e^{-2\alpha^2}\left(b_0b_1\right)\right)^2, P_1^{(I)B\to C} = \left(\frac{N_b}{8N_{1b}}\right)^2\left(\frac{1-e^{-2\alpha^2}}{1+e^{-2\alpha^2}}\right)^2, \quad (14)$$

$$F_1^{(I)C\to A} = \left(N_{1c}N_c\right)^2 e^{\frac{-\pi^2}{8\alpha^2}}\left(\sum_{i=0}^{1} c_i + 2e^{-2\alpha^2}\left(c_0c_1\right)\right)^2, P_1^{(I)C\to A} = \left(\frac{N_c}{8N_{1c}}\right)^2\left(\frac{1-e^{-2\alpha^2}}{1+e^{-2\alpha^2}}\right)^2, \quad (15)$$

$$where, (N_{1a})^{-2} = \left(\sum_{i=0}^{1} a_i^2 - 2e^{-2\alpha^2}\left(a_0a_1\right)\right)^2, (N_{1b})^{-2} = \left(\sum_{i=0}^{1} b_i^2 - 2e^{-2\alpha^2}\left(b_0b_1\right)\right)^2 \ and \ (N_{1c})^{-2} = \left(\sum_{i=0}^{1} c_i^2 - 2e^{-2\alpha^2}\left(c_0c_1\right)\right)^2.$$

Clearly, one may note from Eqs (13-15), in all three cases $A\to B$, $B\to C$ and $C\to A$, 'Near-Faithfull Partial Cyclic QT' results, plotted in Figure 2 to assess dependencies of fidelities on probability-amplitudes $\theta_{1-3}$ as well on $\alpha$, the modal-amplitudes of optical coherent states.

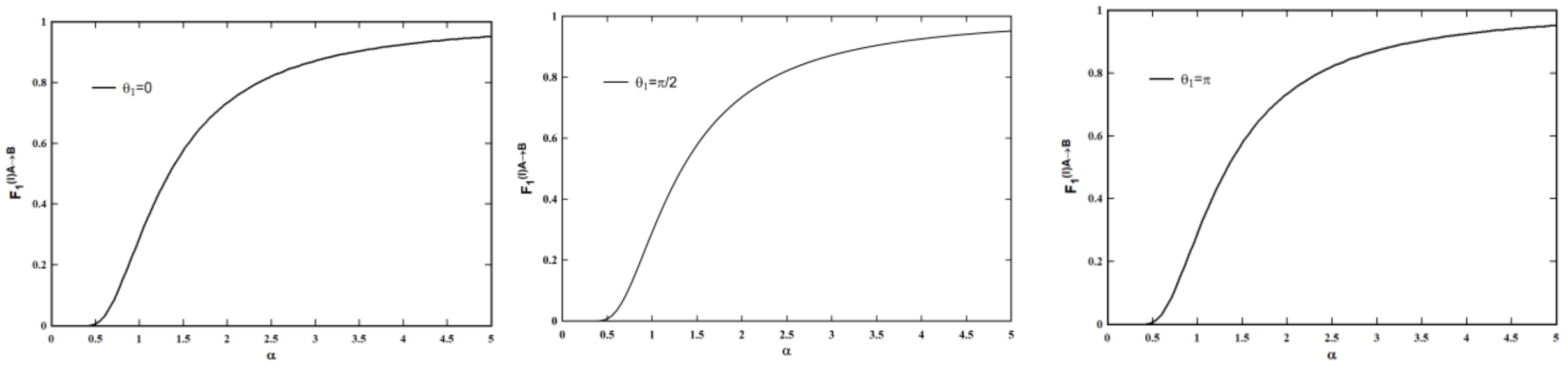

(a) (b) (c)

Figure 2. Fidelity of $F_1^{(I)A\to B}$ vs $\alpha$ for $\Theta_1$= 0(a), $\pi$/2(b), $\pi$(c) . Since expressions for $F_1^{(I)B\to C}$ and $F_1^{(I)C\to A}$ are similar to that of $F_1^{(I)A\to B}$ plots will have similar resemblances.

One may calculate Fidelities for entire Case (I) described by Eqs. (7a). It is seen that Fidelities are ,

$$F_1^{(I)A\to B} = F_2^{(I)A\to B} = F_3^{(I)A\to B} = F_8^{(I)A\to B}; \quad F_4^{(I)A\to B} = F_5^{(I)A\to B} = F_6^{(I)A\to B} = F_7^{(I)A\to B},$$

$$F_1^{(I)B\to C} = F_2^{(I)B\to C} = F_4^{(I)B\to C} = F_6^{(I)B\to C} = F_7^{(I)B\to C}; \quad F_3^{(I)B\to C} = F_5^{(I)B\to C} = F_8^{(I)B\to C},$$

$$F_1^{(I)C\to A} = F_3^{(I)C\to A} = F_4^{(I)C\to A} = F_6^{(I)C\to A}; \quad F_2^{(I)C\to A} = F_5^{(I)C\to A} = F_7^{(I)C\to A} = F_8^{(I)C\to A},$$

Similarly, one may evaluate fidelities for Case (II-VIII), Eqs. (7b-7h),

$$\begin{aligned} & F_1^{(II)A\to B} = F_1^{(III)A\to B} = F_1^{(IV)A\to B} = F_1^{(V)A\to B} = F_1^{(VI)A\to B} = F_1^{(VII)A\to B} = F_1^{(VIII)A\to B} \\ & = F_2^{(II)A\to B} = F_2^{(III)A\to B} = F_2^{(VI)A\to B} = F_2^{(V)A\to B} = F_2^{(VI)A\to B} = F_2^{(VII)A\to B} = F_2^{(VIII)A\to B} \\ & = F_3^{(II)A\to B} = F_3^{(III)A\to B} = F_3^{(IV)A\to B} = F_3^{(V)A\to B} = F_3^{(VI)A\to B} = F_3^{(VII)A\to B} = F_3^{(VIII)A\to B} \\ & = F_8^{(II)A\to B} = F_8^{(III)A\to B} = F_8^{(IV)A\to B} = F_8^{(V)A\to B} = F_8^{(VI)A\to B} = F_8^{(VII)A\to B} = F_8^{(VIII)A\to B}, \end{aligned}$$

$$\begin{aligned} & F_4^{(II)A\to B} = F_4^{(III)A\to B} = F_4^{(IV)A\to B} = F_4^{(V)A\to B} = F_4^{(VI)A\to B} = F_4^{(VII)A\to B} = F_4^{(VIII)A\to B} \\ & = F_5^{(II)A\to B} = F_5^{(III)A\to B} = F_5^{(IV)A\to B} = F_5^{(V)A\to B} = F_5^{(VI)A\to B} = F_5^{(VII)A\to B} = F_5^{(VIII)A\to B} \\ & = F_6^{(II)A\to B} = F_6^{(III)A\to B} = F_6^{(IV)A\to B} = F_6^{(V)A\to B} = F_6^{(VI)A\to B} = F_6^{(VII)A\to B} = F_6^{(VIII)A\to B} \\ & = F_7^{(II)A\to B}, = F_7^{(III)A\to B} = F_7^{(IV)A\to B} = F_7^{(V)A\to B} = F_7^{(VI)A\to B} = F_7^{(VII)A\to B} = F_7^{(VIII)A\to B} \end{aligned}$$

$$\begin{aligned} & F_1^{(II)B\to C} = F_1^{(III)B\to C} = F_1^{(IV)B\to C} = F_1^{(V)B\to C} = F_1^{(VI)B\to C} = F_1^{(VII)B\to C} = F_1^{(VIII)B\to C} \\ & = F_2^{(II)B\to C} = F_2^{(III)B\to C} = F_2^{(IV)B\to C} = F_2^{(V)B\to C} = F_2^{(VI)B\to C} = F_2^{(VII)B\to C} = F_2^{(VIII)B\to C} \\ & = F_4^{(II)B\to C} = F_4^{(III)B\to C} = F_4^{(IV)B\to C} = F_4^{(V)B\to C} = F_4^{(VI)B\to C} = F_4^{(VII)B\to C} = F_4^{(VIII)B\to C} \\ & = F_6^{(II)B\to C} = F_6^{(III)B\to C} = F_6^{(IV)B\to C} = F_6^{(V)B\to C} = F_6^{(VI)B\to C} = F_6^{(VII)B\to C} = F_6^{(VIII)B\to C} \\ & = F_7^{(II)B\to C} = F_7^{(III)B\to C} = F_7^{(IV)B\to C} = F_7^{(V)B\to C} = F_7^{(VI)B\to C} = F_7^{(VII)B\to C} = F_7^{(VIII)B\to C}, \end{aligned}$$

$$
\begin{aligned}
F_3^{(II)B\to C} &= F_3^{(III)B\to C} = F_3^{(IV)B\to C} = F_3^{(V)B\to C} = F_3^{(VI)B\to C} = F_3^{(VII)B\to C} = F_3^{(VIII)B\to C} \\
&= F_3^{(II)B\to C} = F_3^{(III)B\to C} = F_3^{(IV)B\to C} = F_3^{(V)B\to C} = F_3^{(VI)B\to C} = F_3^{(VII)B\to C} = F_3^{(VIII)B\to C} \\
&= F_5^{(II)B\to C} = F_5^{(III)B\to C} = F_5^{(IV)B\to C} = F_5^{(V)B\to C} = F_5^{(VI)B\to C} = F_5^{(VII)B\to C} = F_5^{(VIII)B\to C} \\
&= F_8^{(II)B\to C} = F_8^{(III)B\to C} = F_8^{(IV)B\to C} = F_8^{(V)B\to C} = F_8^{(VI)B\to C} = F_8^{(VII)B\to C} = F_8^{(VIII)B\to C}
\end{aligned}
$$

$$
\begin{aligned}
F_1^{(II)C\to A} &= F_1^{(III)C\to A} = F_1^{(IV)C\to A} = F_1^{(V)C\to A} = F_1^{(VI)C\to A} = F_1^{(VII)C\to A} = F_1^{(VIII)C\to A} \\
&= F_3^{(II)C\to A} = F_3^{(III)C\to A} = F_3^{(IV)C\to A} = F_3^{(V)C\to A} = F_3^{(VI)C\to A} = F_3^{(VII)C\to A} = F_3^{(VIII)C\to A} \\
&= F_4^{(II)C\to A} = F_4^{(III)C\to A} = F_4^{(IV)C\to A} = F_4^{(V)C\to A} = F_4^{(VI)C\to A} = F_4^{(VII)C\to A} = F_4^{(VIII)C\to A} \\
&= F_6^{(II)C\to A} = F_6^{(III)C\to A} = F_6^{(IV)C\to A} = F_6^{(V)C\to A} = F_6^{(VI)C\to A} = F_6^{(VII)C\to A} = F_6^{(VIII)C\to A},
\end{aligned}
$$

$$
\begin{aligned}
F_2^{(II)C\to A} &= F_2^{(III)C\to A} = F_2^{(IV)C\to A} = F_2^{(V)C\to A} = F_2^{(VI)C\to A} = F_2^{(VII)C\to A} = F_2^{(VIII)C\to A} \\
&= F_5^{(II)C\to A} = F_5^{(III)C\to A} = F_5^{(IV)C\to A} = F_5^{(V)C\to A} = F_5^{(VI)C\to A} = F_5^{(VII)C\to A} = F_5^{(VIII)C\to A} \\
&= F_7^{(II)C\to A} = F_7^{(III)C\to A} = F_7^{(IV)C\to A} = F_7^{(V)C\to A} = F_7^{(VI)C\to A} = F_7^{(VII)C\to A} = F_7^{(VIII)C\to A} \\
&= F_8^{(II)C\to A} = F_8^{(III)C\to A} = F_8^{(IV)C\to A} = F_8^{(V)C\to A} = F_8^{(VI)C\to A} = F_8^{(VII)C\to A} = F_8^{(VIII)C\to A}
\end{aligned}
$$

Finally, the 'Average Fidelity' of entire scheme from $(A \to B)$, $(B \to C)$ and $(C \to A)$ can be defined as,

$$F_{av}^{A\to B} = \sum_{i=1}^{8} (F_i^{(I)A\to B} P_i^{(I)A\to B} + F_i^{(II)A\to B} P_i^{(II)A\to B} + \dots\dots\dots + F_i^{(VIII)A\to B} P_i^{(VIII)A\to B}) \tag{16}$$

$$F_{av}^{B\to C} = \sum_{i=1}^{8} (F_i^{(I)B\to C} P_i^{(I)B\to C} + F_i^{(II)B\to C} P_i^{(II)B\to C} + \dots\dots\dots + F_i^{(VIII)B\to C} P_i^{(VIII)B\to C}) \tag{17}$$

$$F_{av}^{C\to A} = \sum_{i=1}^{8} (F_i^{(I)C\to A} P_i^{(I)C\to A} + F_i^{(II)C\to A} P_i^{(II)B\to A} + \dots\dots\dots + F_i^{(VIII)C\to A} P_i^{(VIII)C\to A}) \tag{18}$$

Plots of Eq. (16), Figure 3, clearly, displays that Average fidelity is independent of probability-amplitudes parameters as well as on $\alpha$, modal-amplitude of optical coherent states.

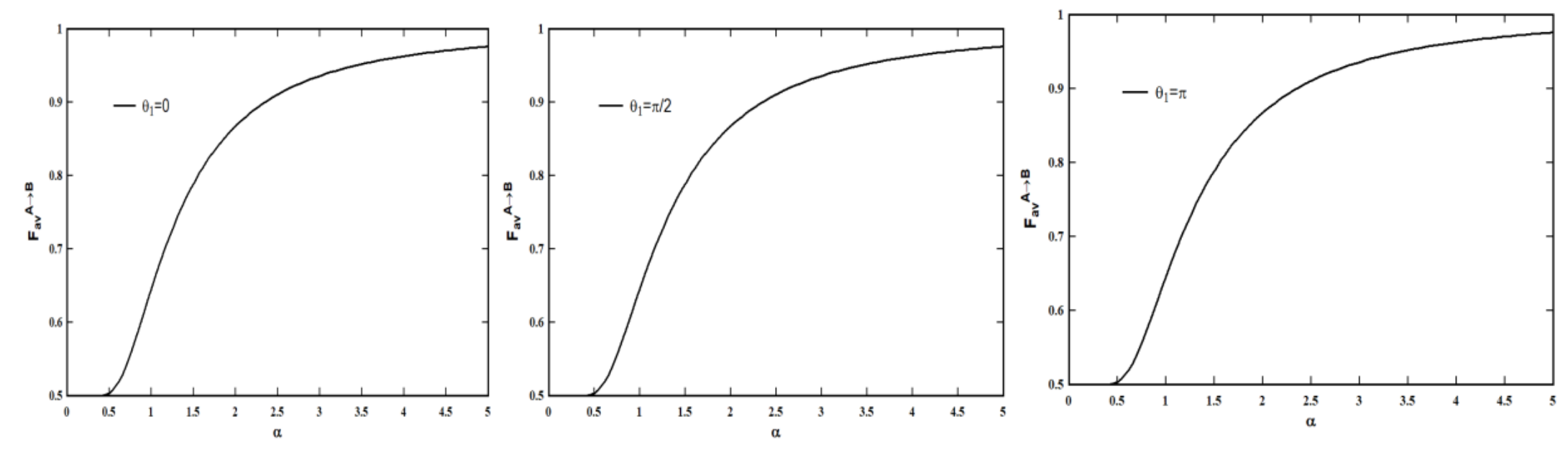

(a) (b) (c)

Figure 3. Fidelity of $F_{av}^{A\to B}$ vs $\alpha$ for $\Theta_1$=0 (a), $\pi/2$ (b), $\pi$ (c). Plots for $F_{av}^{B\to C}$ and $F_{av}^{C\to A}$ will be similar as expressions are equivalent.

## IV. Conclusions and Future Prospects:

We have presented a protocol for a unidirectional simultaneous faithful Cyclic QT. The communication complexity of the protocol may, easily, be recognized as three classical bits: one classical bit from Alice to Bob, one classical bit from Bob to Alice and one classical bit from Alice to Charlie, if one encodes (ODD, EVEN) as (0, 1), respectively, and the three e-bits because three Bell Coherent-states are needed to prepare quantum resource employed as the quantum channel. The protocol for simultaneous unidirectional cyclic quantum teleportation with optical coherent-states shows promise for experimental realization as involved devices are feasible and prevalent in Quantum Optics. However, error-mitigation and fault-tolerance imposes formidable challenges.

**Appendix A:** Mixing various modes of quantum-information states Eqs.(3-5) and those of quantum channel (Eq.1) at 'Symmetric Beam Splitter with phase shifter', one obtains,

$$
\begin{aligned}
&|\phi\rangle_{7,8,9,10,11,12,4,5,6} = N_T[(a_0b_0c_0|\sqrt{2\alpha},0\rangle_{7,8}|\sqrt{2\alpha},0\rangle_{9,10}|\sqrt{2\alpha},0\rangle_{11,12}|\alpha,\alpha,\alpha\rangle_{4,5,6} + |\sqrt{2\alpha},0\rangle_{7,8}|\sqrt{2\alpha},0\rangle_{9,10}|0,\sqrt{2\alpha}\rangle_{11,12}|\alpha,\alpha,-\alpha\rangle_{4,5,6} \\
&+|\sqrt{2\alpha},0\rangle_{7,8}|0,\sqrt{2\alpha}\rangle_{9,10}|\sqrt{2\alpha},0\rangle_{11,12}|\alpha,-\alpha,\alpha\rangle_{4,5,6} + |\sqrt{2\alpha},0\rangle_{7,8}|0,\sqrt{2\alpha}\rangle_{9,10}|0,\sqrt{2\alpha}\rangle_{11,12}|\alpha,-\alpha,-\alpha\rangle_{4,5,6} \\
&+|0,\sqrt{2\alpha}\rangle_{7,8}|\sqrt{2\alpha},0\rangle_{9,10}|\sqrt{2\alpha},0\rangle_{11,12}|-\alpha,\alpha,\alpha\rangle + |0,\sqrt{2\alpha}\rangle_{7,8}|\sqrt{2\alpha},0\rangle_{9,10}|0,\sqrt{2\alpha}\rangle_{11,12}|-\alpha,\alpha,-\alpha\rangle_{4,5,6} \\
&+|0,\sqrt{2\alpha}\rangle_{7,8}|0,\sqrt{2\alpha}\rangle_{9,10}|\sqrt{2\alpha},0\rangle_{11,12}|-\alpha,-\alpha,\alpha\rangle_{4,5,6} + |0,\sqrt{2\alpha}\rangle_{7,8}|0,\sqrt{2\alpha}\rangle_{9,10}|0,\sqrt{2\alpha}\rangle_{11,12}|-\alpha,-\alpha,-\alpha\rangle_{4,5,6}) \\
&+(a_0b_0c_1|\sqrt{2\alpha},0\rangle_{7,8}|\sqrt{2\alpha},0\rangle_{9,10}|0,-\sqrt{2\alpha}\rangle_{11,12}|\alpha,\alpha,\alpha\rangle_{4,5,6} + |\sqrt{2\alpha},0\rangle_{7,8}|\sqrt{2\alpha},0\rangle_{9,10}|-\sqrt{2\alpha},0\rangle_{11,12}|\alpha,\alpha,-\alpha\rangle_{4,5,6} \\
&+|\sqrt{2\alpha},0\rangle_{7,8}|0,\sqrt{2\alpha}\rangle_{9,10}|0,-\sqrt{2\alpha}\rangle_{11,12}|\alpha,-\alpha,\alpha\rangle_{4,5,6} + |\sqrt{2\alpha},0\rangle_{7,8}|0,\sqrt{2\alpha}\rangle_{9,10}|-\sqrt{2\alpha},0\rangle_{11,12}|\alpha,-\alpha,-\alpha\rangle_{4,5,6} \\
&+|0,\sqrt{2\alpha}\rangle_{7,8}|\sqrt{2\alpha},0\rangle_{9,10}|0,-\sqrt{2\alpha}\rangle_{11,12}|-\alpha,\alpha,\alpha\rangle_{4,5,6} + |0,\sqrt{2\alpha}\rangle_{7,8}|\sqrt{2\alpha},0\rangle_{9,10}|-\sqrt{2\alpha},0\rangle_{11,12}|-\alpha,\alpha,-\alpha\rangle_{4,5,6} \\
&+|0,\sqrt{2\alpha}\rangle_{7,8}|0,\sqrt{2\alpha}\rangle_{9,10}|0,-\sqrt{2\alpha}\rangle_{11,12}|-\alpha,-\alpha,\alpha\rangle_{4,5,6} + |0,\sqrt{2\alpha}\rangle_{7,8}|0,\sqrt{2\alpha}\rangle_{9,10}|-\sqrt{2\alpha},0\rangle_{11,12}|-\alpha,-\alpha,-\alpha\rangle_{4,56}) \\
&+(a_0b_1c_0|\sqrt{2\alpha},0\rangle_{7,8}|0,-\sqrt{2\alpha}\rangle_{9,10}|\sqrt{2\alpha},0\rangle_{11,12}|\alpha,\alpha,\alpha\rangle + |\sqrt{2\alpha},0\rangle_{7,8}|0,-\sqrt{2\alpha}\rangle_{9,10}|0,\sqrt{2\alpha}\rangle_{11,12}|\alpha,\alpha,-\alpha\rangle_{4,5,6} \\
&+|\sqrt{2\alpha},0\rangle_{7,8}|-\sqrt{2\alpha},0\rangle_{9,10}|\sqrt{2\alpha},0\rangle_{11,12}|\alpha,-\alpha,\alpha\rangle_{4,5,6} + |\sqrt{2\alpha},0\rangle_{7,8}|-\sqrt{2\alpha},0\rangle_{9,10}|0,\sqrt{2\alpha}\rangle_{11,12}|\alpha,-\alpha,-\alpha\rangle_{4,5,6} \\
&+|0,\sqrt{2\alpha}\rangle_{7,8}|0,-\sqrt{2\alpha}\rangle_{9,10}|\sqrt{2\alpha},0\rangle_{11,12}|-\alpha,\alpha,\alpha\rangle_{4,5,6} + |0,\sqrt{2\alpha}\rangle_{7,8}|0,-\sqrt{2\alpha}\rangle_{9,10}|0,\sqrt{2\alpha}\rangle_{11,12}|-\alpha,\alpha,-\alpha\rangle_{4,5,6} \\
&+|0,\sqrt{2\alpha}\rangle_{7,8}|-\sqrt{2\alpha},0\rangle_{9,10}|\sqrt{2\alpha},0\rangle_{11,12}|-\alpha,-\alpha,\alpha\rangle_{4,5,6} + |0,\sqrt{2\alpha}\rangle_{7,8}|-\sqrt{2\alpha},0\rangle_{9,10}|0,\sqrt{2\alpha}\rangle_{11,12}|-\alpha,-\alpha,-\alpha\rangle_{4,5,6}) \\
&+(a_0b_1c_1|\sqrt{2\alpha},0\rangle_{7,8}|0,-\sqrt{2\alpha}\rangle_{9,10}|0,-\sqrt{2\alpha}\rangle_{11,12}|\alpha,\alpha,\alpha\rangle_{4,5,6} + |\sqrt{2\alpha},0\rangle_{7,8}|0,-\sqrt{2\alpha}\rangle_{9,10}|-\sqrt{2\alpha},0\rangle_{11,12}|\alpha,\alpha,-\alpha\rangle_{4,5,6} \\
&+|\sqrt{2\alpha},0\rangle_{7,8}|-\sqrt{2\alpha},0\rangle_{9,10}|0,-\sqrt{2\alpha}\rangle_{11,12}|\alpha,-\alpha,\alpha\rangle_{4,5,6} + |\sqrt{2\alpha},0\rangle_{7,8}|-\sqrt{2\alpha},0\rangle_{9,10}|-\sqrt{2\alpha},0\rangle_{11,12}|\alpha,-\alpha,-\alpha\rangle_{4,5,6} \\
&+|0,\sqrt{2\alpha}\rangle_{7,8}|0,-\sqrt{2\alpha}\rangle_{9,10}|0,-\sqrt{2\alpha}\rangle_{11,12}|-\alpha,\alpha,\alpha\rangle_{4,5,6} + |0,\sqrt{2\alpha}\rangle_{7,8}|0,-\sqrt{2\alpha}\rangle_{9,10}|-\sqrt{2\alpha},0\rangle_{11,12}|-\alpha,\alpha,-\alpha\rangle_{4,5,6} \\
&+|0,\sqrt{2\alpha}\rangle_{7,8}|-\sqrt{2\alpha},0\rangle_{9,10}|0,-\sqrt{2\alpha}\rangle_{11,12}|-\alpha,-\alpha,\alpha\rangle_{4,5,6} + |0,\sqrt{2\alpha}\rangle_{7,8}|-\sqrt{2\alpha},0\rangle_{9,10}|-\sqrt{2\alpha},0\rangle_{11,12}|-\alpha,-\alpha,-\alpha\rangle_{4,5,6}) \\
&+(a_1b_0c_1|0,-\sqrt{2\alpha},\rangle_{7,8}|\sqrt{2\alpha},0\rangle_{9,10}|0,-\sqrt{2\alpha}\rangle_{11,12}|\alpha,\alpha,\alpha\rangle_{4,5,6} + |0,-\sqrt{2\alpha}\rangle_{7,8}|\sqrt{2\alpha},0\rangle_{9,10}|-\sqrt{2\alpha},0\rangle_{11,12}|\alpha,\alpha,-\alpha\rangle_{4,5,6} \\
&+|0,-\sqrt{2\alpha}\rangle_{7,8}|0,\sqrt{2\alpha},\rangle_{9,10}|0,-\sqrt{2\alpha}\rangle_{11,12}|\alpha,-\alpha,\alpha\rangle_{4,5,6} + |0,-\sqrt{2\alpha}\rangle_{7,8}|0,\sqrt{2\alpha}\rangle_{9,10}|-\sqrt{2\alpha},0\rangle_{11,12}|\alpha,-\alpha,-\alpha\rangle_{4,5,6} \\
&+|-\sqrt{2\alpha},0\rangle_{7,8}|\sqrt{2\alpha},0\rangle_{9,10}|0,-\sqrt{2\alpha}\rangle_{11,12}|-\alpha,\alpha,\alpha\rangle_{4,5,6} + |-\sqrt{2\alpha},0\rangle_{7,8}|\sqrt{2\alpha},0\rangle_{9,10}|-\sqrt{2\alpha},0\rangle_{11,12}|-\alpha,\alpha,-\alpha\rangle_{4,5,6} \\
&+|-\sqrt{2\alpha},0\rangle_{7,8}|0,\sqrt{2\alpha}\rangle_{9,10}|0,-\sqrt{2\alpha}\rangle_{11,12}|-\alpha,-\alpha,\alpha\rangle_{4,5,6} + |-\sqrt{2\alpha},0\rangle_{7,8}|0,\sqrt{2\alpha}\rangle_{9,10}|-\sqrt{2\alpha},0\rangle_{11,12}|-\alpha,-\alpha,-\alpha\rangle_{4,5,6}) \\
&+(a_1b_0c_1|0,-\sqrt{2\alpha},\rangle_{7,8}|\sqrt{2\alpha},0\rangle_{9,10}|0,-\sqrt{2\alpha}\rangle_{11,12}|\alpha,\alpha,\alpha\rangle_{4,5,6} + |0,-\sqrt{2\alpha}\rangle_{7,8}|\sqrt{2\alpha},0\rangle_{9,10}|-\sqrt{2\alpha},0\rangle_{11,12}|\alpha,\alpha,-\alpha\rangle_{4,5,6} \\
&+|0,-\sqrt{2\alpha}\rangle_{7,8}|0,\sqrt{2\alpha},\rangle_{9,10}|0,-\sqrt{2\alpha}\rangle_{11,12}|\alpha,-\alpha,\alpha\rangle_{4,5,6} + |0,-\sqrt{2\alpha}\rangle_{7,8}|0,\sqrt{2\alpha}\rangle_{9,10}|-\sqrt{2\alpha},0\rangle_{11,12}|\alpha,-\alpha,-\alpha\rangle_{4,5,6} \\
&+|-\sqrt{2\alpha},0\rangle_{7,8}|\sqrt{2\alpha},0\rangle_{9,10}|0,-\sqrt{2\alpha}\rangle_{11,12}|-\alpha,\alpha,\alpha\rangle_{4,5,6} + |-\sqrt{2\alpha},0\rangle_{7,8}|\sqrt{2\alpha},0\rangle_{9,10}|-\sqrt{2\alpha},0\rangle_{11,12}|-\alpha,\alpha,-\alpha\rangle_{4,5,6} \\
&+|-\sqrt{2\alpha},0\rangle_{7,8}|0,\sqrt{2\alpha}\rangle_{9,10}|0,-\sqrt{2\alpha}\rangle_{11,12}|-\alpha,-\alpha,\alpha\rangle_{4,5,6} + |-\sqrt{2\alpha},0\rangle_{7,8}|0,\sqrt{2\alpha}\rangle_{9,10}|-\sqrt{2\alpha},0\rangle_{11,12}|-\alpha,-\alpha,-\alpha\rangle_{4,5,6}) \\
&+(a_1b_1c_0|0,-\sqrt{2\alpha},\rangle_{7,8}|0,-\sqrt{2\alpha},\rangle_{9,10}|\sqrt{2\alpha},0\rangle_{11,12}|\alpha,\alpha,\alpha\rangle_{4,5,6} + |0,-\sqrt{2\alpha}\rangle_{7,8}|0,-\sqrt{2\alpha}\rangle_{9,10}|0,\sqrt{2\alpha}\rangle_{11,12}|\alpha,\alpha,-\alpha\rangle_{4,5,6} \\
&+|0,-\sqrt{2\alpha}\rangle_{7,8}|-\sqrt{2\alpha},0\rangle_{9,10}|\sqrt{2\alpha},0\rangle_{11,12}|\alpha,-\alpha,\alpha\rangle_{4,5,6} + |0,-\sqrt{2\alpha}\rangle_{7,8}|-\sqrt{2\alpha},\rangle_{9,10}|0,\sqrt{2\alpha}\rangle_{11,12}|\alpha,-\alpha,-\alpha\rangle_{4,5,6} \\
&+|-\sqrt{2\alpha},0\rangle_{7,8}|0,-\sqrt{2\alpha},0\rangle_{9,10}|\sqrt{2\alpha},0\rangle_{11,12}|-\alpha,\alpha,\alpha\rangle_{4,5,6} + |-\sqrt{2\alpha},0\rangle_{7,8}|0,-\sqrt{2\alpha}\rangle_{9,10}|0,\sqrt{2\alpha},\rangle_{11,12}|-\alpha,\alpha,-\alpha\rangle_{4,5,6} \\
&+|-\sqrt{2\alpha},0\rangle_{7,8}|-\sqrt{2\alpha},0\rangle_{9,10}|\sqrt{2\alpha},0\rangle_{11,12}|-\alpha,-\alpha,\alpha\rangle_{4,5,6} + |-\sqrt{2\alpha},0\rangle_{7,8}|-\sqrt{2\alpha},0\rangle_{9,10}|0,\sqrt{2\alpha},\rangle_{11,12}|-\alpha,-\alpha,-\alpha\rangle_{4,5,6}) \\
&+(a_1b_1c_1|0,-\sqrt{2\alpha},\rangle_{7,8}|0,-\sqrt{2\alpha},\rangle_{9,10}|0,-\sqrt{2\alpha},\rangle_{11,12}|\alpha,\alpha,\alpha\rangle_{4,5,6} + |0,-\sqrt{2\alpha}\rangle_{7,8}|0,-\sqrt{2\alpha}\rangle_{9,10}|-\sqrt{2\alpha},0\rangle_{11,12}|\alpha,\alpha,-\alpha\rangle_{4,5,6} \\
&+|0,-\sqrt{2\alpha}\rangle_{7,8}|-\sqrt{2\alpha},0\rangle_{9,10}|0,-\sqrt{2\alpha},\rangle_{11,12}|\alpha,-\alpha,\alpha\rangle_{4,5,6} + |0,-\sqrt{2\alpha}\rangle_{7,8}|-\sqrt{2\alpha},\rangle_{9,10}|-\sqrt{2\alpha},0\rangle_{11,12}|\alpha,-\alpha,-\alpha\rangle_{4,5,6} \\
&+|-\sqrt{2\alpha},0\rangle_{7,8}|0,-\sqrt{2\alpha},0\rangle_{9,10}|0,-\sqrt{2\alpha},\rangle_{11,12}|-\alpha,\alpha,\alpha\rangle_{4,5,6} + |-\sqrt{2\alpha},0\rangle_{7,8}|0,-\sqrt{2\alpha}\rangle_{9,10}|-\sqrt{2\alpha},0\rangle_{11,12}|-\alpha,\alpha,-\alpha\rangle_{4,5,6} \\
&+|-\sqrt{2\alpha},0\rangle_{7,8}|-\sqrt{2\alpha},0\rangle_{9,10}|0,-\sqrt{2\alpha},\rangle_{11,12}|-\alpha,-\alpha,\alpha\rangle_{4,5,6} + |-\sqrt{2\alpha},0\rangle_{7,8}|-\sqrt{2\alpha},0\rangle_{9,10}|-\sqrt{2\alpha},0\rangle_{11,12}|-\alpha,-\alpha,-\alpha\rangle_{4,5,6})]
\end{aligned}
$$

**(A1)**

**Appendix B**

| No of photon detected in Alice, Bob and Charlie Lab. | | | Collapse State corresponding to $n_7, n_9, n_{11} \neq 0$ $n_8, n_{10}, n_{12} = 0$ (see Eq. (7a)) | Unitary Operation | | | Collapse State corresponding to $n_8, n_{10}, n_{12} \neq 0$ $n_7, n_9, n_{11} = 0$ (see Eq. (7b)) | Unitary Operation | | | Faithful(F) /Near Faithful (NF) |
|---|---|---|---|---|---|---|---|---|---|---|---|
| Alice $n_7/\ n_8$ | Bob $n_9/\ n_{10}$ | Charlie $n_{11}/n_{12}$ | | Alice | Bob | Charlie | | Alice | Bob | Charlie | |
| Odd | Odd | Odd | $N_a(a_0\|\alpha\rangle - a_1\|-\alpha\rangle)_4 \otimes$<br>$N_b(b_0\|\alpha\rangle - b_1\|-\alpha\rangle)_5 \otimes$<br>$N_c(c_0\|\alpha\rangle - c_1\|-\alpha\rangle)_6$ | $D_6$ | $D_4$ | $D_5$ | $N_a(a_0\|-\alpha\rangle - a_1\|\alpha\rangle)_4 \otimes$<br>$N_b(b_0\|-\alpha\rangle - b_1\|\alpha\rangle)_5 \otimes$<br>$N_c(c_0\|-\alpha\rangle - c_1\|\alpha\rangle)_6$ | $D_6 \otimes P_6$ | $D_4 \otimes P_4$ | $D_5 \otimes P_5$ | NF |
| Odd | Odd | Even | $N_a(a_0\|\alpha\rangle - a_1\|-\alpha\rangle)_4 \otimes$<br>$N_b(b_0\|\alpha\rangle - b_1\|-\alpha\rangle)_5 \otimes$<br>$N_c(c_0\|\alpha\rangle + c_1\|-\alpha\rangle)_6$ | $I_6$ | $D_4$ | $D_5$ | $N_a(a_0\|-\alpha\rangle + a_1\|\alpha\rangle)_4 \otimes$<br>$N_b(b_0\|-\alpha\rangle + b_1\|\alpha\rangle)_5 \otimes$<br>$N_c(c_0\|-\alpha\rangle - c_1\|\alpha\rangle)_6$ | $P_6$ | $D_4 \otimes P_4$ | $D_5 \otimes P_5$ | NF |
| Odd | Even | Odd | $N_a(a_0\|\alpha\rangle - a_1\|-\alpha\rangle)_4 \otimes$<br>$N_b(b_0\|\alpha\rangle + b_1\|-\alpha\rangle)_5 \otimes$<br>$N_c(c_0\|\alpha\rangle - c_1\|-\alpha\rangle)_6$ | $D_6$ | $D_4$ | $I_5$ | $N_a(a_0\|-\alpha\rangle + a_1\|\alpha\rangle)_4 \otimes$<br>$N_b(b_0\|-\alpha\rangle - b_1\|\alpha\rangle)_5 \otimes$<br>$N_c(c_0\|-\alpha\rangle + c_1\|\alpha\rangle)_6$ | $D_6 \otimes P_6$ | $D_4 \otimes P_4$ | $P_5$ | NF |
| Even | Odd | Odd | $N_a(a_0\|\alpha\rangle + a_1\|-\alpha\rangle)_4 \otimes$<br>$N_b(b_0\|\alpha\rangle - b_1\|-\alpha\rangle)_5 \otimes$<br>$N_c(c_0\|\alpha\rangle - c_1\|-\alpha\rangle)_6$ | $D_6$ | $I_4$ | $D_5$ | $N_a(a_0\|-\alpha\rangle - a_1\|\alpha\rangle)_4 \otimes$<br>$N_b(b_0\|-\alpha\rangle + b_1\|\alpha\rangle)_5 \otimes$<br>$N_c(c_0\|-\alpha\rangle + c_1\|\alpha\rangle)_6$ | $D_6 \otimes P_6$ | $P_4$ | $D_5 \otimes P_5$ | NF |
| Even | Even | Even | $N_a(a_0\|\alpha\rangle + a_1\|-\alpha\rangle)_4 \otimes$<br>$N_b(b_0\|\alpha\rangle + b_1\|-\alpha\rangle)_5 \otimes$<br>$N_c(c_0\|\alpha\rangle + c_1\|-\alpha\rangle)_6$ | $I_6$ | $I_4$ | $I_5$ | $N_a(a_0\|-\alpha\rangle + a_1\|\alpha\rangle)_4 \otimes$<br>$N_b(b_0\|-\alpha\rangle + b_1\|\alpha\rangle)_5 \otimes$<br>$N_c(c_0\|-\alpha\rangle + c_1\|\alpha\rangle)_6$ | $P_6$ | $P_4$ | $P_5$ | F |
| Even | Even | Odd | $N_a(a_0\|\alpha\rangle + a_1\|-\alpha\rangle)_4 \otimes$<br>$N_b(b_0\|\alpha\rangle - b_1\|-\alpha\rangle)_5 \otimes$<br>$N_c(c_0\|\alpha\rangle - c_1\|-\alpha\rangle)_6$ | $D_6$ | $I_4$ | $D_5$ | $N_a(a_0\|-\alpha\rangle + a_1\|\alpha\rangle)_4 \otimes$<br>$N_b(b_0\|-\alpha\rangle + b_1\|\alpha\rangle)_5 \otimes$<br>$N_c(c_0\|-\alpha\rangle - c_1\|\alpha\rangle)_6$ | $D_6 \otimes P_6$ | $P_4$ | $D_5 \otimes P_5$ | NF |
| Even | Odd | Even | $N_a(a_0\|\alpha\rangle + a_1\|-\alpha\rangle)_4 \otimes$<br>$N_b(b_0\|\alpha\rangle - b_1\|-\alpha\rangle)_5 \otimes$<br>$N_c(c_0\|\alpha\rangle + c_1\|-\alpha\rangle)_6$ | $I_6$ | $I_4$ | $D_5$ | $N_a(a_0\|-\alpha\rangle + a_1\|\alpha\rangle)_4 \otimes$<br>$N_b(b_0\|-\alpha\rangle - b_1\|\alpha\rangle)_5 \otimes$<br>$N_c(c_0\|-\alpha\rangle + c_1\|\alpha\rangle)_6$ | $P_6$ | $P_4$ | $D_5 \otimes P_5$ | NF |
| Odd | Even | Even | $N_a(a_0\|\alpha\rangle - a_1\|-\alpha\rangle)_4 \otimes$<br>$N_b(b_0\|\alpha\rangle + b_1\|-\alpha\rangle)_5 \otimes$<br>$N_c(c_0\|\alpha\rangle + c_1\|-\alpha\rangle)_6$ | $I_6$ | $D_4$ | $I_5$ | $N_a(a_0\|-\alpha\rangle - a_1\|\alpha\rangle)_4 \otimes$<br>$N_b(b_0\|-\alpha\rangle + b_1\|\alpha\rangle)_5 \otimes$<br>$N_c(c_0\|-\alpha\rangle + c_1\|\alpha\rangle)_6$ | $P_6$ | $D_4 \otimes P_4$ | $P_5$ | NF |

Table I. All possible detection-events for cases-I and II ($n_7, n_9, n_{11} \neq 0$, $n_8, n_{10}, n_{12} = 0$ and $n_8, n_{10}, n_{12} \neq 0$ $n_7, n_9, n_{11} = 0$) and corresponding necessary unitary operator for the successful cyclic QT protocol.

| No of photon detected in Alice, Bob and Charlie Lab. | | | Collapse State corresponding to $n_7, n_9, n_{12} \neq 0$ $n_8, n_{10}, n_{11} = 0$ (see Eq. (7c)) | Unitary Operation | | | Collapse State corresponding to $n_7, n_{10}, n_{11} = 0$ $n_8, n_9, n_{12} = 0$ (see Eq. (7d)) | Unitary Operation | | | Faithful(F) /Near Faithful (NF) |
|---|---|---|---|---|---|---|---|---|---|---|---|
| Alice $n_7/n_8$ | Bob $n_9/n_{10}$ | Charlie $n_{11}/n_{12}$ | | Alice | Bob | Charlie | | Alice | Bob | Charlie | |
| Odd | Odd | Odd | $N_a(a_0\|\alpha\rangle - a_1\|-\alpha\rangle)_4 \otimes$<br>$N_b(b_0\|\alpha\rangle - b_1\|-\alpha\rangle)_5 \otimes$<br>$N_c(c_0\|-\alpha\rangle - c_1\|\alpha\rangle)_6$ | $D_6 \otimes P_6$ | $D_4$ | $D_5$ | $N_a(a_0\|\alpha\rangle - a_1\|-\alpha\rangle)_4 \otimes$<br>$N_b(b_0\|-\alpha\rangle - b_1\|\alpha\rangle)_5 \otimes$<br>$N_c(c_0\|\alpha\rangle - c_1\|-\alpha\rangle)_6$ | $D_6$ | $D_4$ | $D_5 \otimes P_5$ | NF |
| Odd | Odd | Even | $N_a(a_0\|\alpha\rangle - a_1\|-\alpha\rangle)_4 \otimes$<br>$N_b(b_0\|\alpha\rangle - b_1\|-\alpha\rangle)_5 \otimes$<br>$N_c(c_0\|-\alpha\rangle + c_1\|\alpha\rangle)_6$ | $P_6$ | $D_4$ | $D_5$ | $N_a(a_0\|\alpha\rangle - a_1\|-\alpha\rangle)_4 \otimes$<br>$N_b(b_0\|-\alpha\rangle - b_1\|\alpha\rangle)_5 \otimes$<br>$N_c(c_0\|\alpha\rangle + c_1\|-\alpha\rangle)_6$ | $I_6$ | $D_4$ | $D_5 \otimes P_5$ | NF |
| Odd | Even | Odd | $N_a(a_0\|\alpha\rangle - a_1\|-\alpha\rangle)_4 \otimes$<br>$N_b(b_0\|\alpha\rangle + b_1\|-\alpha\rangle)_5 \otimes$<br>$N_c(c_0\|-\alpha\rangle - c_1\|\alpha\rangle)_6$ | $D_6 \otimes P_6$ | $D_4$ | $I_5$ | $N_a(a_0\|\alpha\rangle - a_1\|-\alpha\rangle)_4 \otimes$<br>$N_b(b_0\|-\alpha\rangle + b_1\|\alpha\rangle)_5 \otimes$<br>$N_c(c_0\|\alpha\rangle - c_1\|-\alpha\rangle)_6$ | $D_6$ | $D_4$ | $P_5$ | NF |
| Even | Odd | Odd | $N_a(a_0\|\alpha\rangle + a_1\|-\alpha\rangle)_4 \otimes$<br>$N_b(b_0\|\alpha\rangle - b_1\|-\alpha\rangle)_5 \otimes$<br>$N_c(c_0\|-\alpha\rangle - c_1\|\alpha\rangle)_6$ | $D_6 \otimes P_6$ | $I_4$ | $D_5$ | $N_a(a_0\|\alpha\rangle + a_1\|-\alpha\rangle)_4 \otimes$<br>$N_b(b_0\|-\alpha\rangle - b_1\|-\rangle)_5 \otimes$<br>$N_c(c_0\|\alpha\rangle - c_1\|-\alpha\rangle)_6$ | $D_6$ | $I_4$ | $D_5 \otimes P_5$ | NF |
| Even | Even | Even | $N_a(a_0\|\alpha\rangle + a_1\|-\alpha\rangle)_4 \otimes$<br>$N_b(b_0\|\alpha\rangle + b_1\|-\alpha\rangle)_5 \otimes$<br>$N_c(c_0\|-\alpha\rangle + c_1\|\alpha\rangle)_6$ | $P_6$ | $I_4$ | $I_5$ | $N_a(a_0\|\alpha\rangle + a_1\|-\alpha\rangle)_4 \otimes$<br>$N_b(b_0\|-\alpha\rangle + b_1\|\alpha\rangle)_5 \otimes$<br>$N_c(c_0\|\alpha\rangle + c_1\|-\alpha\rangle)_6$ | $I_6$ | $I_4$ | $P_5$ | F |
| Even | Even | Odd | $N_a(a_0\|\alpha\rangle + a_1\|-\alpha\rangle)_4 \otimes$<br>$N_b(b_0\|\alpha\rangle - b_1\|-\alpha\rangle)_5 \otimes$<br>$N_c(c_0\|-\alpha\rangle - c_1\|\alpha\rangle)_6$ | $D_6 \otimes P_6$ | $I_4$ | $D_5$ | $N_a(a_0\|\alpha\rangle + a_1\|-\alpha\rangle)_4 \otimes$<br>$N_b(b_0\|-\alpha\rangle - b_1\|\alpha\rangle)_5 \otimes$<br>$N_c(c_0\|\alpha\rangle - c_1\|-\alpha\rangle)_6$ | $D_6$ | $I_4$ | $D_5 \otimes P_5$ | NF |
| Even | Odd | Even | $N_a(a_0\|\alpha\rangle + a_1\|-\alpha\rangle)_4 \otimes$<br>$N_b(b_0\|\alpha\rangle - b_1\|-\alpha\rangle)_5 \otimes$<br>$N_c(c_0\|-\alpha\rangle + c_1\|\alpha\rangle)_6$ | $P_6$ | $I_4$ | $D_5$ | $N_a(a_0\|\alpha\rangle + a_1\|-\alpha\rangle)_4 \otimes$<br>$N_b(b_0\|-\alpha\rangle - b_1\|\alpha\rangle)_5 \otimes$<br>$N_c(c_0\|\alpha\rangle + c_1\|-\alpha\rangle)_6$ | $I_6$ | $I_4$ | $D_5 \otimes P_5$ | NF |
| Odd | Even | Even | $N_a(a_0\|\alpha\rangle - a_1\|-\alpha\rangle)_4 \otimes$<br>$N_b(b_0\|\alpha\rangle + b_1\|-\alpha\rangle)_5 \otimes$<br>$N_c(c_0\|-\alpha\rangle + c_1\|\alpha\rangle)_6$ | $P_6$ | $D_4$ | $I_5$ | $N_a(a_0\|\alpha\rangle - a_1\|-\alpha\rangle)_4 \otimes$<br>$N_b(b_0\|-\alpha\rangle + b_1\|\alpha\rangle)_5 \otimes$<br>$N_c(c_0\|\alpha\rangle + c_1\|-\alpha\rangle)_6$ | $I_6$ | $D_4$ | $P_5$ | NF |

Table II. All possible detection-events for cases-III and IV ($n_7, n_9, n_{12} \neq 0$, $n_8, n_{10}, n_{11} = 0$ and $n_7, n_{10}, n_{11} \neq 0$ $n_8, n_9, n_{12} = 0$) and corresponding necessary unitary operator for the successful cyclic QT protocol.

| No of photon detected in Alice, Bob and Charlie Lab. | | | Collapse State corresponding to $n_7, n_{10}, n_{12} \neq 0$ $n_8, n_9, n_{11} = 0$ (see Eq. (7e)) | Unitary Operation | | | Collapse State corresponding to $n_8, n_9, n_{11} \neq 0$ $n_7, n_{10}, n_{12} = 0$ (see Eq. (7f)) | Unitary Operation | | | Faithful(F) /Near Faithful (NF) |
|---|---|---|---|---|---|---|---|---|---|---|---|
| Alice $n_7/ n_8$ | Bob $n_9$ / $n_{10}$ | Charlie $n_{11}/n_{12}$ | | Alice | Bob | Charlie | | Alice | Bob | Charlie | |
| Odd | Odd | Odd | $N_a(a_0\|\alpha\rangle - a_1\|-\alpha\rangle)_4 \otimes$ $N_b(b_0\|-\alpha\rangle - b_1\|\alpha\rangle)_5 \otimes$ $N_c(c_0\|-\alpha\rangle - c_1\|\alpha\rangle)_6$ | $D_6 \otimes P_6$ | $D_4$ | $D_5 \otimes P_5$ | $N_a(a_0\|-\alpha\rangle - a_1\|\alpha\rangle)_4 \otimes$ $N_b(b_0\|\alpha\rangle - b_1\|-\alpha\rangle)_5 \otimes$ $N_c(c_0\|\alpha\rangle - c_1\|-\alpha\rangle)_6$ | $D_6$ | $D_4 \otimes P_4$ | $D_5$ | NF |
| Odd | Odd | Even | $N_a(a_0\|\alpha\rangle - a_1\|-\alpha\rangle)_4 \otimes$ $N_b(b_0\|-\alpha\rangle - b_1\|\alpha\rangle)_5 \otimes$ $N_c(c_0\|-\alpha\rangle + c_1\|\alpha\rangle)_6$ | $P_6$ | $D_4$ | $D_5 \otimes P_5$ | $N_a(a_0\|-\alpha\rangle - a_1\|\alpha\rangle)_4 \otimes$ $N_b(b_0\|\alpha\rangle - b_1\|-\alpha\rangle)_5 \otimes$ $N_c(c_0\|\alpha\rangle + c_1\|-\alpha\rangle)_6$ | $I_6$ | $D_4 \otimes P_4$ | $D_5$ | NF |
| Odd | Even | Odd | $N_a(a_0\|\alpha\rangle - a_1\|-\alpha\rangle)_4 \otimes$ $N_b(b_0\|-\alpha\rangle + b_1\|\alpha\rangle)_5 \otimes$ $N_c(c_0\|-\alpha\rangle - c_1\|\alpha\rangle)_6$ | $D_6 \otimes P_6$ | $D_4$ | $P_5$ | $N_a(a_0\|-\alpha\rangle - a_1\|\alpha\rangle)_4 \otimes$ $N_b(b_0\|\alpha\rangle + b_1\|-\alpha\rangle)_5 \otimes$ $N_c(c_0\|\alpha\rangle - c_1\|-\alpha\rangle)_6$ | $D_6$ | $D_4 \otimes P_4$ | $I_5$ | NF |
| Even | Odd | Odd | $N_a(a_0\|\alpha\rangle + a_1\|-\alpha\rangle)_4 \otimes$ $N_b(b_0\|-\alpha\rangle - b_1\|\alpha\rangle)_5 \otimes$ $N_c(c_0\|-\alpha\rangle - c_1\|\alpha\rangle)_6$ | $D_6 \otimes P_6$ | $I_4$ | $D_5 \otimes P_5$ | $N_a(a_0\|-\alpha\rangle + a_1\|\alpha\rangle)_4 \otimes$ $N_b(b_0\|\alpha\rangle - b_1\|-\alpha\rangle)_5 \otimes$ $N_c(c_0\|\alpha\rangle - c_1\|-\alpha\rangle)_6$ | $D_6$ | $P_4$ | $D_5$ | NF |
| Even | Even | Even | $N_a(a_0\|\alpha\rangle + a_1\|-\alpha\rangle)_4 \otimes$ $N_b(b_0\|-\alpha\rangle + b_1\|\alpha\rangle)_5 \otimes$ $N_c(c_0\|-\alpha\rangle + c_1\|\alpha\rangle)_6$ | $P_6$ | $I_4$ | $P_5$ | $N_a(a_0\|-\alpha\rangle + a_1\|\alpha\rangle)_4 \otimes$ $N_b(b_0\|\alpha\rangle + b_1\|-\alpha\rangle)_5 \otimes$ $N_c(c_0\|\alpha\rangle + c_1\|-\alpha\rangle)_6$ | $I_6$ | $P_4$ | $I_5$ | F |
| Even | Even | Odd | $N_a(a_0\|\alpha\rangle + a_1\|-\alpha\rangle)_4 \otimes$ $N_b(b_0\|-\alpha\rangle - b_1\|\alpha\rangle)_5 \otimes$ $N_c(c_0\|-\alpha\rangle - c_1\|\alpha\rangle)_6$ | $D_6 \otimes P_6$ | $I_4$ | $D_5 \otimes P_5$ | $N_a(a_0\|-\alpha\rangle + a_1\|\alpha\rangle)_4 \otimes$ $N_b(b_0\|\alpha\rangle - b_1\|-\alpha\rangle)_5 \otimes$ $N_c(c_0\|\alpha\rangle - c_1\|-\alpha\rangle)_6$ | $D_6$ | $P_4$ | $D_5$ | NF |
| Even | Odd | Even | $N_a(a_0\|\alpha\rangle + a_1\|-\alpha\rangle)_4 \otimes$ $N_b(b_0\|-\alpha\rangle - b_1\|\alpha\rangle)_5 \otimes$ $N_c(c_0\|-\alpha\rangle + c_1\|\alpha\rangle)_6$ | $P_6$ | $I_4$ | $D_5 \otimes P_5$ | $N_a(a_0\|-\alpha\rangle + a_1\|\alpha\rangle)_4 \otimes$ $N_b(b_0\|\alpha\rangle - b_1\|-\alpha\rangle)_5 \otimes$ $N_c(c_0\|\alpha\rangle + c_1\|-\alpha\rangle)_6$ | $I_6$ | $P_4$ | $D_5$ | NF |
| Odd | Even | Even | $N_a(a_0\|\alpha\rangle - a_1\|-\alpha\rangle)_4 \otimes$ $N_b(b_0\|-\alpha\rangle + b_1\|\alpha\rangle)_5 \otimes$ $N_c(c_0\|-\alpha\rangle + c_1\|\alpha\rangle)_6$ | $P_6$ | $D_4$ | $P_5$ | $N_a(a_0\|-\alpha\rangle - a_1\|\alpha\rangle)_4 \otimes$ $N_b(b_0\|\alpha\rangle + b_1\|-\alpha\rangle)_5 \otimes$ $N_c(c_0\|\alpha\rangle + c_1\|-\alpha\rangle)_6$ | $I_6$ | $D_4 \otimes P_4$ | $I_5$ | NF |

Table III. All detection-events for cases-V and VI ($n_7, n_{10}, n_{12} \neq 0$, $n_8, n_9, n_{11} = 0$ and $n_8, n_9, n_{11} \neq 0$ $n_7, n_{10}, n_{12} = 0$) and corresponding necessary unitary operator for the successful cyclic QT protocol.

| No of photon detected in Alice, Bob and Charlie Lab. | | | Collapse State corresponding to $n_8, n_9, n_{12} \neq 0$ $n_7, n_{10}, n_{11} = 0$ (see Eq. (7g)) | Unitary Operation | | | Collapse State corresponding to $n_8, n_{10}, n_{12} \neq 0$ $n_7, n_9, n_{11} = 0$ (see Eq. (7h)) | Unitary Operation | | | Faithful(F) /Near Faithful (NF) |
|---|---|---|---|---|---|---|---|---|---|---|---|
| Alice $n_7/\, n_8$ | Bob $n_9/$ $n_{10}$ | Charlie $n_{11}/n_{12}$ | | Alice | Bob | Charlie | | Alice | Bob | Charlie | |
| Odd | Odd | Odd | $N_a(a_0\|-\alpha\rangle - a_1\|\alpha\rangle)_4 \otimes$ $N_b(b_0\|\alpha\rangle - b_1\|-\alpha\rangle)_5 \otimes$ $N_c(c_0\|-\alpha\rangle - c_1\|\alpha\rangle)_6$ | $D_6 \otimes P_6$ | $D_4 \otimes P_4$ | $D_5$ | $N_a(a_0\|-\alpha\rangle - a_1\|\alpha\rangle)_4 \otimes$ $N_b(a_0\|-\alpha\rangle - a_1\|\alpha\rangle)_5 \otimes$ $N_c(a_0\|\alpha\rangle - a_1\|-\alpha\rangle)_6$ | $D_6$ | $D_4 \otimes P_4$ | $D_5 \otimes P_5$ | NF |
| Odd | Odd | Even | $N_a(a_0\|-\alpha\rangle - a_1\|\alpha\rangle)_4 \otimes$ $N_b(b_0\|\alpha\rangle - b_1\|-\alpha\rangle)_5 \otimes$ $N_c(c_0\|-\alpha\rangle + c_1\|\alpha\rangle)_6$ | $P_6$ | $D_4 \otimes P_4$ | $D_5$ | $N_a(a_0\|-\alpha\rangle + a_1\|\alpha\rangle)_4 \otimes$ $N_b(a_0\|-\alpha\rangle + a_1\|\alpha\rangle)_5 \otimes$ $N_c(a_0\|\alpha\rangle - a_1\|-\alpha\rangle)_6$ | $D_6$ | $P_4$ | $P_5$ | NF |
| Odd | Even | Odd | $N_a(a_0\|-\alpha\rangle - a_1\|\alpha\rangle)_4 \otimes$ $N_b(b_0\|\alpha\rangle + b_1\|-\alpha\rangle)_5 \otimes$ $N_c(c_0\|-\alpha\rangle - c_1\|\alpha\rangle)_6$ | $D_6 \otimes P_6$ | $D_4 \otimes P_4$ | $I_5$ | $N_a(a_0\|-\alpha\rangle + a_1\|\alpha\rangle)_4 \otimes$ $N_b(a_0\|-\alpha\rangle - a_1\|\alpha\rangle)_5 \otimes$ $N_c(a_0\|\alpha\rangle + a_1\|-\alpha\rangle)_6$ | $I_6$ | $P_4$ | $D_5 \otimes P_5$ | NF |
| Even | Odd | Odd | $N_a(a_0\|-\alpha\rangle + a_1\|\alpha\rangle)_4 \otimes$ $N_b(b_0\|\alpha\rangle - b_1\|-\alpha\rangle)_5 \otimes$ $N_c(c_0\|-\alpha\rangle - c_1\|\alpha\rangle)_6$ | $D_6 \otimes P_6$ | $P_4$ | $D_5$ | $N_a(a_0\|-\alpha\rangle - a_1\|\alpha\rangle)_4 \otimes$ $N_b(a_0\|-\alpha\rangle + a_1\|\alpha\rangle)_5 \otimes$ $N_c(a_0\|\alpha\rangle + a_1\|-\alpha\rangle)_6$ | $I_6$ | $D_4 \otimes P_4$ | $P_5$ | NF |
| Even | Even | Even | $N_a(a_0\|-\alpha\rangle + a_1\|\alpha\rangle)_4 \otimes$ $N_b(b_0\|\alpha\rangle + b_1\|-\alpha\rangle)_5 \otimes$ $N_c(c_0\|-\alpha\rangle + c_1\|\alpha\rangle)_6$ | $P_6$ | $P_4$ | $I_5$ | $N_a(a_0\|-\alpha\rangle + a_1\|\alpha\rangle)_4 \otimes$ $N_b(a_0\|-\alpha\rangle + a_1\|\alpha\rangle)_5 \otimes$ $N_c(a_0\|\alpha\rangle + a_1\|-\alpha\rangle)_6$ | $I_6$ | $P_4$ | $P_5$ | F |
| Even | Even | Odd | $N_a(a_0\|-\alpha\rangle + a_1\|\alpha\rangle)_4 \otimes$ $N_b(b_0\|\alpha\rangle - b_1\|-\alpha\rangle)_5 \otimes$ $N_c(c_0\|-\alpha\rangle - c_1\|\alpha\rangle)_6$ | $D_6 \otimes P_6$ | $P_4$ | $D_5$ | $N_a(a_0\|-\alpha\rangle + a_1\|\alpha\rangle)_4 \otimes$ $N_b(a_0\|-\alpha\rangle + a_1\|\alpha\rangle)_5 \otimes$ $N_c(a_0\|\alpha\rangle - a_1\|-\alpha\rangle)_6$ | $D_6$ | $P_4$ | $P_5$ | NF |
| Even | Odd | Even | $N_a(a_0\|-\alpha\rangle + a_1\|\alpha\rangle)_4 \otimes$ $N_b(b_0\|\alpha\rangle - b_1\|-\alpha\rangle)_5 \otimes$ $N_c(c_0\|-\alpha\rangle + c_1\|\alpha\rangle)_6$ | $P_6$ | $P_4$ | $D_5$ | $N_a(a_0\|-\alpha\rangle + a_1\|\alpha\rangle)_4 \otimes$ $N_b(a_0\|-\alpha\rangle - a_1\|\alpha\rangle)_5 \otimes$ $N_c(a_0\|\alpha\rangle + a_1\|-\alpha\rangle)_6$ | $I_6$ | $P_4$ | $D_5 \otimes P_5$ | NF |
| Odd | Even | Even | $N_a(a_0\|-\alpha\rangle - a_1\|\alpha\rangle)_4 \otimes$ $N_b(b_0\|\alpha\rangle + b_1\|-\alpha\rangle)_5 \otimes$ $N_c(c_0\|-\alpha\rangle + c_1\|\alpha\rangle)_6$ | $P_6$ | $D_4 \otimes P_4$ | $I_5$ | $N_a(a_0\|-\alpha\rangle - a_1\|\alpha\rangle)_4 \otimes$ $N_b(a_0\|-\alpha\rangle + a_1\|\alpha\rangle)_5 \otimes$ $N_c(a_0\|\alpha\rangle + a_1\|-\alpha\rangle)_6$ | $I_6$ | $D_4 \otimes P_4$ | $P_5$ | NF |

Table IV. All possible detection-events for cases-VII and VIII ($n_8, n_9, n_{12} \neq 0, n_7, n_{10}, n_{11} = 0$and $n_8, n_{10}, n_{11} \neq 0$ $n_7, n_9, n_{12} = 0$) and corresponding necessary unitary operator for the successful cyclic QT protocol.